# Comment on: "Estimating the Hartree—Fock limit from finite basis set calculations" [Jensen F (2005) Theor Chem Acc 113:267]

Amir Karton and Jan M.L. Martin

*Department of Organic Chemistry, Weizmann Institute of Science, 76100 Rehovot, Israel*
*Email: comartin@wicc.weizmann.ac.il*

**Abstract**

We demonstrate that a minor modification of the extrapolation proposed by Jensen [(2005): Theor Chem Acc 113:267] yields very reliable estimates of the Hartree-Fock limit in conjunction with correlation consistent basis sets. Specifically, a two-point extrapolation of the form $E_{HF,L} = E_{HF,\infty} + A(L+1)\exp(-9\sqrt{L})$ yields HF limits $E_{HF,\infty}$ with an RMS error of 0.1 millihartree using aug-cc-pVQZ and aug-cc-pV5Z basis sets, and of 0.01 millihartree using aug-cc-pV5Z and aug-cc-pV6Z basis sets.

*Keywords:* Basis set convergence, Hartree-Fock limit, extrapolation formulas, correlation-consistent basis sets

**Introduction**

In a recent paper,[1] Jensen considers the estimation of the Hartree-Fock limit from finite basis set calculations, using a recently computed database[2] of numerical Hartree-Fock energies of diatomic molecules as a reference. He concludes: (a) that finite basis sets, particularly his own polarization consistent (pc-n) basis sets,[3] can approach the Hartree-Fock limit very closely; (b) that their limiting convergence is very well described by an extrapolation formula of the form

$$E(L) = E_\infty + A(L+1)\exp\left(-B\sqrt{n_s}\right) \qquad (1)$$

where L is the highest angular momentum present in the basis set and $n_s$ the number of s-type basis functions for the heaviest atom present.

Jensen also found that this formula works rather less well for the correlation consistent (cc) basis sets of Dunning and coworkers,[4,5] which are commonly used in high-accuracy ab initio thermochemistry calculations (e.g., in W1/W2/W3 theory[6] and in the HEAT project[7]). He ascribed this both to the intrinsic nature of the cc basis sets — which were optimized for recovering correlation energy — and to the fact that $n_s$ does not change smoothly in the cc-pVnZ series.

Normally, however, $n_s$ will be roughly proportional to $L$ in a sequence of basis sets with increasing L. We will show below that substitution of $L$ for $n_s$ leads to an SCF extrapolation formula that yields highly satisfactory SCF limits from aug-cc-pVnZ basis sets, provided these are of at least aug-cc-pVQZ quality.

**Methods**

All calculations were carried out using MOLPRO[8] running on the Martin group's Linux cluster at the Weizmann Institute of Science. All energies were converged to at least 10 decimal places, and integral screening thresholds adjusted accordingly. Restricted open-shell Hartree-Fock was used for the open-shell species.

We carried out SCF calculations for the molecules in Ref.2, at the reference geometries given there, using the aug-cc-pVnZ and aug-cc-pV(n+d)Z basis sets (n=D,T,Q,5,6) on first- and second-row atoms, respectively. (In the remainder of the article, the notation AVnZ is used for brevity.) These latter basis sets[5] contain an extra high-exponent d function for second-row atoms, in order to recover "inner polarization" effects.[9] For a subset of molecules, we additionally carried out calculations with the aug-pV7Z basis sets of Valeev et al.:[10] for $S_2$ and SO, we added an additional high-exponent d function by multiplying the highest d already present by a factor of 2.5. The AV(6+d)Z and AV7Z basis sets were obtained from the EMSL Basis Set Library.[11]

**Results and discussion**

A complete sets of total energies can be obtained from the authors, while a summary of error statistics and parameters can be found in Table 1.

The raw AVnZ RMS (root mean square) errors decay from 2 millihartree with the AVQZ basis set to 316 and 57 microhartree, respectively, for the AV5Z and AV6Z basis sets. The uncontracted aug-pc3 and aug-pc4 basis sets display RMS errors of 358 and 81 microhartree, respectively — one should keep in mind these actually have $L_{max}$ a step lower than AVnZ basis sets of comparable quality.

The $A+B/L^5$ extrapolation used in W1 theory performs quite well from the AV{T,Q}Z basis set pair actually used there (339 microhartree, comparable to the AV5Z basis set) but clearly overshoots the mark for larger basis sets — the errors of 574 and 118 microhartree, respectively, for the AV{Q,5}Z and AV{5,6}Z pairs actually exceed those of the raw AV5Z and AV6Z results, respectively. For the the aug-pc{3,4} basis sets, the $A+B/L^5$ extrapolation results in an RMS error of 64 microhartree, not much better than the raw aug-pc4 result.

Three-point geometric extrapolation with the AV{T,Q,5}Z basis sets yields an RMS error of 344 microhartree, which is inferior to the raw AV5Z data (although it errs on the opposite side from $A+B/L^5$). The same extrapolation from AV{Q,5,6}Z data reduces error to 27 microhartree, which does represent an improvement on the raw data. Using the equation $A + B(L+1)\exp(-\gamma\sqrt{L})$ instead as a three-point extrapolation, we find an RMS error of 41 microhartree (and an average γ of 8.11).

Let us now consider empirical two-point extrapolations, first of the form $A+B/L^\alpha$, where α is an effective decay exponent obtained by minimizing the RMS error. From the AV{T,Q}Z pair we obtain α=5.34 for a respectably small RMS error of 206 microhartree. For the AV{Q,5}Z pair α=8.74 with an RMS error of 128 microhartree; for the AV{5,6}Z pair (with α=9.43) this can be brought down to a paltry 10 microhartree — basically exact for all thermochemical purposes.

Two-point extrapolations of the forms

$$E(L) = E_\infty + A\exp(-\beta\sqrt{L}) \quad (2a)$$
$$E(L) = E_\infty + A(L+1)\exp(-\gamma\sqrt{L}) \quad (2b)$$

(with β and γ again global parameters) will obviously result in the same RMS errors, as all such extrapolations can be written in the form

$$E_\infty = E(L_2) + \frac{E(L_2) - E(L_1)}{c_1 - 1}$$
$$E_\infty = E(L_2) + \frac{E(L_2) - E(L_1)}{c_1 - 1} \quad (3)$$

(with $c_1$ a constant). In fact, the various exponents (α for inverse power law, β for geometric/exponential, and γ for the exponential-√ L form) can easily be shown to be related through the following equalities:

$$\ln c_1 = \alpha \cdot \ln\frac{L_2}{L_1} = \beta \cdot (L_2 - L_1) = \gamma \cdot (\sqrt{L_2} - \sqrt{L_1}) - \ln\frac{L_2+1}{L_1+1} \quad (4)$$

What concerns us here is the values of the effective exponents: for the geometric/exponential form β=1.95 for the AV{Q,5}Z pair, and β=1.72 for the AV{5,6}Z pair. For the exponential-√L form, however, the change in γ is quite small: from 9.03 for the AV{Q,5}Z pair to 8.77 for the AV{5,6}Z pair, a relative change of only 3%. In fact, using γ=9 for *both* cases only increases the error from 9.91 to 10.42 for the AV{5,6}Z pair, and from 128.48 to 128.51 for the AV{Q,5}Z pair.

Optimizing for the combined RMS error $(\text{RMSD}_{56}^2 + \text{RMSD}_{Q5}^2)^{1/2}$ leads to γ=9.0247, which changes to γ=8.9591 if we weigh the two contributions by the inverses of their optimal RMS error (which gives much greater weight to the AV{5,6}Z pair). The average of both exponents is γ=8.992, not significantly different from γ=9. We thus propose as an SCF extrapolation formula for the aug-cc-pV(n+d)Z basis sets (n≥Q):

$$E = E_\infty + A(L+1)\exp(-9\sqrt{L}) \quad (5)$$

which implies

$$E_\infty = E_L + \frac{E_L - E_{L-1}}{\dfrac{L\exp\left(9(\sqrt{L}-\sqrt{L-1})\right)}{L+1} - 1} \quad (6)$$

This expression, sadly, does not work very well for the AV{T,Q}Z pair. We considered a number of alternative expressions, such as A+B(L+1)exp(-γ'√ (L+D)), A+B(L+1)exp(-γ''√ L)+C(L+1)exp(γ''L), and the like, to no avail. It appears that the AVTZ basis set simply does not lend itself well to SCF extrapolation.

What about the uncontracted aug-pc*n* basis sets? Two-point extrapolation from aug-pc3 and aug-pc4 basis sets yields an RMS error of only 24 microhartree (for an optimal exponent γ=7.27), while a respectable 88 microhartree is obtained using the comparatively small aug-pc2 and aug-pc3 basis sets (for an optimal γ=11.79).

Does Eq.(5) hold tolerably well for basis sets larger than AV6Z? If it does, then it should give good predictions of the SCF/aug-cc-pV(7+d)Z energy. We have calculated the latter for the following ten molecules: $C_2$, $CN^-$, $CO$, $F_2$, $FH$, $N_2$, $NO^+$, $O_2$, $S_2$, and $SO$. (The k functions were omitted for technical reasons: consideration of the f, g, h and i function contributions for the AV7Z basis sets — which decay by about an order of magnitude apiece — suggests the k function contribution to the SCF energy is an order of magnitude below the microhartree range for the first-row species, but that contributions on the order of 1-2 microhartrees cannot be ruled out for the second-row species.)

RMS deviation between predicted and actual AV7Z SCF energies is 5.4 microhartree, suggesting that our equation works tolerably well even in that region. A two-point AV{6,7}Z extrapolation with an optimized α parameter ($\alpha=8.18$, $\gamma=7.10$) yields an RMS error of only 3.5 microhartree for the SCF limits, compared to 15.4 microhartree for the raw AV7Z energies. However, inspection of the individual errors reveals that the two second-row species, $S_2$ and SO, are outliers (possibly because of neglect of k function contributions to the AV7Z energies). Upon eliminating them and reoptimizing, we find an RMSD of only 1.5 microhartree ($\alpha=9.06$, $\gamma=7.80$), compared to 10.3 microhartree for raw AV7Z. For the first-row species, Eq.(5) has an RMSD of just 3.1 microhartree.

Finally, we note that our effective decay exponents for the large basis sets are considerably smaller than those previously obtained by Schwenke[12] for a much smaller sample of systems. His study was primarily concerned with basis set convergence of the *correlation* energy, and his SCF energies are indeed converged to his stated goal of 10 microhartree. However, numerical Hartree-Fock calculations with the 2DHF program[13] at his reference geometries revealed that his estimated HF limits were on average about 10 microhartree above the true limits[14] — quite enough to affect the effective exponents for AV{Q,5}Z and AV{5,6}Z basis set pairs.

**Conclusions**

While the polarization consistent basis sets are extremely valuable in Hartree-Fock and DFT calculations, it is quite possible to obtain reliable Hartree-Fock limits from SCF energies obtained with (sufficiently large) correlation consistent basis sets using Eqs. (4,5). This is particularly relevant for accurate ab initio computational thermochemistry work, where these basis sets are commonly used because of their efficiency in recovering valence correlation. We recommend that eq. (5) be used instead of $A+B/L^5$ in W2 and W3 calculations; for W1, $A+B/L^5$ is close to optimal.

**Acknowledgments**

This work was supported by the Lise Meitner-Minerva Center for Computational Quantum Chemistry (of which JMLM is a member *ad personam*) and by the Helen and Martin Kimmel Center for Molecular Design. JMLM is the incumbent of the Baroness Thatcher Professorial Chair of Chemistry.

Table 1: Summary of RMS deviations (microhartree) from the numerical Hartree-Fock energies and parameters for various extrapolation formulas. $L_{max}$ represents the highest angular momentum in any basis set involved.

|  | $A+BL^{-\alpha}$ | $A+B\exp(-\beta L)$ | $A+B(L+1)\cdot \exp(-\gamma\sqrt{L})$ | RMSD | $L_{max}$ |
|---|---|---|---|---|---|
|  | $\alpha$ | $\beta$ | $\gamma$ |  |  |
| raw AVQZ | — | — | — | 2032 | 4 |
| raw AV5Z | — | — | — | 316 | 5 |
| raw AV6Z | — | — | — | 57 | 6 |
| raw aug-pc3uncon | — | — | — | 358 | 4 |
| raw aug-pc4uncon | — | — | — | 81 | 5 |
| AV{T,Q}Z | [5] | — | — | 339 | 4 |
| AV{Q,5}Z | [5] | — | — | 574 | 5 |
| AV{5,6}Z | [5] | — | — | 118 | 6 |
| aug-pc{3,4}uncon | [5] | — | — | 64 | 5 |
| AV{T,Q,5}Z | — | variable | — | 344 | 5 |
| AV{Q,5,6}Z | — | variable | — | 27 | 6 |
| AV{Q,5,6}Z | — | — | variable | 41 | 6 |
| aug-pc{2,3,4}uncon | — | variable | — | 67 | 5 |
| AV{T,Q}Z | 5.34 | 1.54 | 6.57 | 206 | 4 |
| AV{Q,5}Z | 8.74 | 1.95 | 9.03 | 128 | 5 |
| AV{5,6}Z | 9.43 | 1.72 | 8.77 | 10 | 6 |
| aug-pc{2,3}uncon | 10.21 | 2.94 | 11.79 | 88 | 4 |
| aug-pc{3,4}uncon | 6.87 | 1.53 | 7.27 | 24 | 5 |
| AV{Q,5}Z | — | — | [9] | 128.51 | 5 |
| AV{5,6}Z | — | — | [9] | 10.42 | 6 |
| raw AV7Z | — | — | — | 15.4(10.3[a]) | 7 |
| AV{5,6,7}Z | — | variable | — | 8.6(5.1[a]) | 7 |
| AV{5,6,7}Z | — | — | variable | 7.1(4.1[a]) | 7 |
| AV{6,7}Z | 8.18 | 1.26 | 7.10 | 3.5 | 7 |
| (a) | 9.06[a] | 1.40[a] | 7.80[a] | 1.5[a] | 7 |
| AV{6,7}Z | — | — | [9] | 6.8(3.1[a]) | 7 |

The notation AV{Q,5}Z means a two-point extrapolation from the AVQZ and AV5Z basis sets, while AV{5,6,7}Z refers to three-point extrapolation from AV5Z, AV6Z, and AV7Z basis sets.

A fixed parameter $\alpha$, $\beta$, or $\gamma$ is indicated by square brackets. An entry marked "variable" means that the relevant parameter is determined for every individual case in a 3-point extrapolation.

Error statistics involving the AV7Z basis set only refer to 10 molecules (see text): all other error statistics refer to the complete set of molecules in Ref.2.

(a) First-row species only